\documentclass[showpacs,preprintnumbers,twocolumn,amsmath,amssymb,groupedaddress,superscriptaddress,footinbib]{revtex4}

\usepackage{graphicx,hyperref}
\usepackage{dcolumn}
\usepackage{bm}

\begin{document}

\title{Microscopic analysis of the giant monopole resonance excitation energy}

\author{M. K. Gaidarov}
\affiliation{Institute for Nuclear Research and Nuclear Energy,
Bulgarian Academy of Sciences, Sofia 1784, Bulgaria}

\affiliation{Department of Physics, Faculty of Mathematics and
Natural Sciences, South-West University "Neofit Rilski",
Blagoevgrad, Bulgaria}

\author{M. V. Ivanov}
\affiliation{Institute for Nuclear Research and Nuclear Energy,
Bulgarian Academy of Sciences, Sofia 1784, Bulgaria}

\affiliation{Department of Physics, Faculty of Mathematics and
Natural Sciences, South-West University "Neofit Rilski",
Blagoevgrad, Bulgaria}

\author{Y. I. Katsarov}
\affiliation{Institute for Nuclear Research and Nuclear Energy,
Bulgarian Academy of Sciences, Sofia 1784, Bulgaria}

\author{A. N. Antonov}
\affiliation{Institute for Nuclear Research and Nuclear Energy,
Bulgarian Academy of Sciences, Sofia 1784, Bulgaria}

\author{I. C. Danchev}
\affiliation{Department of Physical and Mathematical Sciences,
School of Arts and Sciences, University of Mount Olive, 652 R.B.
Butler Dr., Mount Olive, NC 28365, USA}

\begin{abstract}
A systematic study of the isoscalar giant monopole resonance
(ISGMR) in a wide range of nuclei from various isotopic chains is
performed within the microscopic self-consistent Skyrme HF+BCS
method and coherent density fluctuation model (CDFM). The
calculations for the nuclear incompressibility are based on the
Brueckner and Barcelona-Catania-Paris-Madrid (BCPM) energy density
functionals for nuclear matter using the capability of the CDFM to
make a transition to the corresponding incompressibility in finite
nuclei. The results obtained by applying of different definitions
of the ISGMR energy, as well as the two energy-density
functionals, are analyzed and compared with the available
experimental data. The consideration includes the peculiarities of
the proton and neutron density distributions and their
corresponding linear size characteristics. In general, a
connection with the measured neutron skin thicknesses is proposed
as a possible way for realistic estimations of the energy of
ISGMR.
\end{abstract}


\maketitle

\section{Introduction}
\label{sec:intro}

The great interest to study the giant resonances both
experimentally and theoretically is provoked by the fundamental
understanding of the nonequilibrium properties of nuclei and the
nuclear force. Particularly, the isoscalar giant monopole
resonance (ISGMR) measures the collective response of the nucleus
to density fluctuations. The energy of this resonance is connected
to the incompressibility of the nucleus, which, in turn, can be
linked to the incompressibility of the infinite nuclear matter,
which is an important ingredient of the nuclear matter equation of
state (EOS). The EOS plays significant role in the description of
heavy-ion nuclear collision \cite{Bertsch88}, the collapse of the
heavy stars in super novae explosions \cite{Bethe79,Oertel2017}
and some astrophysical quantities, such as radii and masses of
neutron stars \cite{Lindblom92}. The 20\% uncertainty of the
currently accepted value of the incompressibility of nuclear
matter is largely driven by the poor determination of the EOS
isospin asymmetry  term \cite{Garg2011}. To improve upon the
precision of this term, experimental measurements of isoscalar
monopole modes are being carried out on isotopic chains, with an
extension from the nuclei on the valley of stability towards
exotic nuclei with large proton-neutron asymmetry. On the other
side, theoretical estimations of the giant monopole excitation
energies have been made for nuclei in the super heavy region for
$Z=114$ and 120, which are predicted by several models as the next
proton magic numbers beyond $Z=82$ \cite{Biswal2014}.

An important primary motivation for studying the ISGMR is to probe
bulk nuclear properties of the nuclear EOS. As such, it is highly
unexpected that effects arising from microscopic shell structure
would appreciably influence the collective behavior of the nucleus
undergoing these excitations. The isoscalar resonances are excited
through low-momentum transfer reactions in inverse kinematics,
that require special detection devices. At present, promising
results have been obtained using active targets. Different
measurements have been conducted on Ni isotopes far from
stability, namely $^{56}$Ni~\cite{Monrozeau2008, Bagchi2015} and
$^{68}$Ni~\cite{Vandebrouck2014, Vandebrouck2015}. In particular,
the experiment with $^{68}$Ni is the first measurement of the
isoscalar monopole response in a short-lived neutron-rich nucleus
using inelastic alpha scattering. The ISGMR was found to be
fragmented, with a possible indication for the soft monopole
resonance. Complementary to using inelastic scattering of
$\alpha$-particles, which has been used to great effect over the
last several decades, in Ref.~\cite{Arroyo2024} the use of
$^{6}$Li as a probe to study the ISGMR in several stable nuclei
($^{58}$Ni, $^{90}$Zr, $^{116}$Sn, and $^{208}$Pb) was explored.
The measured inelastic scattering spectra have shown to agree very
well with the previously measured ISGMR responses from
$\alpha$-particle scattering, thus manifesting the feasibility of
employing $^{6}$Li inelastic scattering in investigations of the
ISGMR.

The discussion on how to extract the incompressibility of nuclear
matter from the ISGMR dates back to the years
1980s~\cite{Bohigas79,Blaizot80} (see also more recent
review~\cite{Garg2018}). The measurement of the centroid energy of
the ISGMR~\cite{Li2010, Patel2012, Blaizot76, Button2017,
Howard2019, Howard2020a, Howard2020b} provides a very sensitive
method to determine the incompressibility value. Theoretical
investigations in various models~\cite{Brueckner70, Shlomo2001,
Chen2012, Anders2013, Su2018, Colo2020, Bonasera2021} with grouped
values of the nuclear matter incompressibility predict different
ISGMR energies. The comparison with the experimental excitation
energies of the ISGMR in finite nuclei could give the constraint
on the nuclear matter incompressibility modulus.

In the present work the incompressibility and the centroid energy
of ISGMR are studied {\it in a wide range of finite nuclei} on the
basis of the Brueckner \cite{Brueckner68, Brueckner69} as well as
BCPM \cite{Baldo2023,Baldo2013,Baldo2017} functionals for nuclear
matter and using the coherent density fluctuation model (CDFM)
(e.g., Refs.~\cite{Ant80, AHP}). The latter is a natural extension
of the Fermi gas model based on the generator coordinate
method~\cite{AHP, Grif57} and includes long-range correlations of
collective type. In our previous
works~\cite{Ant80,AHP,Gaidarov2011, Gaidarov2012, Gaidarov2014,
Antonov2016, Antonov2018, Danchev2020, Gaidarov2020,
Gaidarov2020ch} we have demonstrated the capability of CDFM to be
applied as an alternative way to make a transition from the
properties of nuclear matter to those of finite nuclei
investigating the ISGMR, the nuclear symmetry energy (NSE), the
neutron pressure, and the asymmetric compressibility in finite
nuclei. While there is enough collected information for these key
EOS parameters (although the uncertainty of their determination is
still large), the volume and surface symmetry energies have been
poorly investigated till now. Therefore, in
Ref.~\cite{Gaidarov2021} we proposed a new alternative approach to
calculate the ratio of the surface to volume components of the NSE
in the framework of the CDFM. We have demonstrated that the new
scheme leads to more realistic values that agree better with the
empirical data and exhibits conceptual and operational advantages.

A first attempt to study the ISGMR in Ni, Sn, and Pb isotopes
within the microscopic self-consistent Skyrme HF+BCS method and
CDFM was performed in our previous work~\cite{Gaidarov2023}. The
calculations were based on the Brueckner energy-density functional
(EDF) for nuclear matter. The obtained results have demonstrated
the relevance of the proposed theoretical approach to probe the
excitation energy of the ISGMR in various nuclei. On the other
hand, however, it became clear that another, more refined method
for calculations of the nuclear compressibility has to be used, as
well as more realistic EDFs to be employed in the calculations of
the monopole excitation energy for wider spectrum of nuclei.

In the present work, we perform calculations using an extended
method to calculate the incompressibility and ISGMR energy in
nuclei, in which EDFs of Brueckner {\it et al.}~\cite{Brueckner68,
Brueckner69} and BCPM \cite{Baldo2023,Baldo2013,Baldo2017} are
adopted. The results for the excitation energies of ISGMR are
obtained by using its two definitions given in literature. Apart
from these complimentary considerations, in this work a variety of
different even-even nuclei from $^{40}$Ca to $^{208}$Pb is
considered and results for the energy of the ISGMR are presented
and discussed in comparison with the experimental data. In
addition, we analyze the values of the centroid energies of Sn
isotopes ($A=112-124$), as well as $^{48}$Ca and $^{208}$Pb
nuclei, studying their isotopic sensitivity by adding results on
the base of experimentally measured neutron skin thickness.

The structure of this paper is the following. In
Section~\ref{sec:formalism} we present the theoretical scheme that
includes the common definitions of the excitation energy of ISGMR
and properties of nuclear matter characterizing its density
dependence around normal nuclear matter density, as well as the
CDFM formalism, which provides a way to calculate starting from
nuclear matter quantities the corresponding ones in finite nuclei.
Some of the relationships are given in Appendix A.
Section~\ref{sec:results} contains the numerical results and
discussion. The summary and main conclusions of the study are
given in Section~\ref{sec:conclusions}.

\section{Theoretical scheme}\label{sec:formalism}

\subsection{Excitation energy of the ISGMR}\label{subsec:ISGMR}

In this subsection we present the relationships used to obtain the
energy of the ISGMR through the incompressibility $K^{A}$ of a
nucleus with $Z$ protons and $N$ neutrons. Basically there are two
definitions which will be used further in our calculations. One of
them is (see Ref.~\cite{Brueckner70})
\begin{equation}
E_{ISGMR}=\frac{\hbar}{r_{0}A^{1/3}}\sqrt{\frac{K^{A}}{m}},
\label{eq:1}
\end{equation}
where $A=Z+N$ is the mass number, $m$ is the nucleon mass and $r_{0}$ is a radius parameter deduced from the equilibrium density.

The second definition is given in the scaling method~\cite{Stringari82} (see, e.g., Refs.~\cite{Patel2012, Blaizot76}) as
\begin{equation}
E_{ISGMR}=\hbar \sqrt{\frac{K^{A}}{m<r^{2}>}},
\label{eq:2}
\end{equation}
where $<r^{2}>$ is the mean square mass radius of the nucleus in the ground state.

We note that Eq.~(\ref{eq:2}) does not contain a fit parameter and
directly uses the radii obtained from mean-field calculations. The
results obtained by the usage of Eqs.~(\ref{eq:1})
and~(\ref{eq:2}) will be shown and discussed in
Section~\ref{sec:results}. Also, the variations related with
parameter $r_{0}$ for different nuclei and the comparison with
some results obtained by using $<r^{2}>$ will be presented there.

\subsection{The relation between the key EOS parameters in nuclear matter and finite nuclei in the CDFM} \label{subsec:CDFM}

First, we will give briefly the main relationships of the CDFM
which was suggested and developed in Refs.~\cite{Ant80, AHP}). For
the purposes of the present work we should mention our recent
papers~\cite{Gaidarov2012, Danchev2020, Gaidarov2021} in which
CDFM was intensively used. In it the one-body density matrix
(OBDM) $\rho(\mathbf{r},\mathbf{r}^{\prime})$ of the nucleus is a
superposition of OBDM's $\rho_{x}({\bf r},{\bf r^{\prime}})$ of
spherical ``pieces'' of nuclear matter (called ``fluctons'') with
radius $x$ in which all A nucleons are uniformly distributed:
\begin{equation}
\rho({\bf r},{\bf r^{\prime}})=\int_{0}^{\infty}dx |F(x)|^{2} \rho_{x}({\bf r},{\bf r^{\prime}})\label{eq:3}
\end{equation}
with
\begin{eqnarray}
\rho_{x}({\bf r},{\bf r^{\prime}})&=&3\rho_{0}(x)
\frac{j_{1}(k_{F}(x)|{\bf r}-{\bf r^{\prime}}|)}{(k_{F}(x)|{\bf
r}-{\bf r^{\prime}}|)}\nonumber \\  & \times & \Theta \left
(x-\frac{|{\bf r}+{\bf r^{\prime}}|}{2}\right ),
\label{eq:4}
\end{eqnarray}
where
\begin{eqnarray}
\Theta(y)=\left\{
\begin{array}{c}
  1, y\geq 0\\
  0, y<0 \\
\end{array}
\right.
\label{eq:4a}
\end{eqnarray}
is the step-function of Heaviside. In Eq.~(\ref{eq:4})
\begin{equation}
\rho_{0}(x)=\frac{3A}{4\pi x^{3}}, \label{eq:5}
\end{equation}
\begin{equation}
k_{F}(x)=\left(\frac{3\pi^{2}}{2}\rho_{0}(x)\right )^{1/3}\equiv \frac{\alpha}{x} \label{eq:6}
\end{equation}
with
\begin{equation}
\alpha=\left(\frac{9\pi A}{8}\right )^{1/3}\simeq 1.52A^{1/3} \label{eq:7}
\end{equation}
and $j_{1}$ is the first-order spherical Bessel function.

The nucleon density distribution is given by the diagonal elements of the OBDM~(\ref{eq:3}):
\begin{equation}
\rho({\bf r})=\rho({\bf r},{\bf r})=\int_{0}^{\infty}dx |F(x)|^{2}\rho_{0}(x)\Theta(x-|{\bf r}|).\label{eq:8}
\end{equation}
It can be seen from Eq.~(\ref{eq:8}) that in the case of monotonically decreasing local density ($d\rho/dr\leq 0$) the weight function $|F(x)|^{2}$ in CDFM can be obtained:
\begin{equation}
|F(x)|^{2}=-\frac{1}{\rho_{0}(x)} \left. \frac{d\rho(r)}{dr}\right |_{r=x} . \label{eq:9}
\end{equation}
It is normalized as
\begin{equation}
\int_{0}^{\infty}dx |F(x)|^{2}=1. \label{eq:10}
\end{equation}

Second, we give the main relationships for the symmetry energy of
nuclear matter (NM) $S(\rho)$, that is related to the energy
density  $\varepsilon(\rho_p({\bf r}),\rho_n({\bf r}))$, the
latter depending on proton and neutron densities, and the
asymmetry
$$
\delta = \frac{\rho_n - \rho_p}{\rho_n + \rho_p}.
$$

The expansion of $\varepsilon$ in $\delta$ up to the quadratic term in asymmetric NM has the form:
\begin{equation}
\varepsilon(\rho,\delta) = \varepsilon_0(\rho,\delta=0)+\rho  S(\rho) \delta^2,  \label{eq:11}
\end{equation}
where $\varepsilon_0(\rho,\delta=0)$ is the energy density of symmetric NM. The second term contains the symmetry energy of NM:
\begin{equation}
S(\rho) = \frac{1}{2} \left.\frac{\partial^2 \Big(\frac{\varepsilon(\rho,\delta)}{\rho}\Big)}{\partial\delta^2}\right\vert_{\delta=0} \label{eq:12}
\end{equation}
(see, e.g.~\cite{Gaidarov2021, Diep2003, Chen2011,Colo2014}). Here we remind the relation of $\varepsilon(\rho,\delta)$ to the energy per particle ${E}/{A}$: $$\frac{\varepsilon(\rho,\delta)}{\rho} = \frac{E(\rho,\delta)}{A}.$$

In the CDFM scheme, the symmetry energy of \emph{finite nuclei} has the form:
\begin{equation}
S^A=\int\limits_{0}^{\infty}dx|F(x)|^2S(\rho_0(x)),
\label{eq:13}
\end{equation}
where the weight function $|F(x)|^2$ is given by Eqs.~(\ref{eq:9}) and~(\ref{eq:10}).

Next, we consider the incompressibility in \emph{finite nuclei} $K^A$, which in the CDFM is written in the form
\begin{equation}
K^A=\int\limits_{0}^{\infty}dx|F(x)|^2K(\rho_0(x)).
\label{eq:14}
\end{equation}
In Eq.~(\ref{eq:14}) the weight function $|F(x)|^2$ is obtained using Eqs.~(\ref{eq:9}) and~(\ref{eq:10}), similarly as in Eq.~(\ref{eq:13}).

The function $K(\rho_0(x))$ in the integrand of Eq.~(\ref{eq:14})
is the incompressibility in infinite nuclear matter. The latter
has the form (see, e.g. \cite{Colo2014}):
\begin{multline}\label{eq:15}
K(\rho_0(x))=K_{NM}(\rho_0(x))+K_{\tau}(\rho_0(x))\delta^2\\+K_{Coul}(\rho_0(x))\dfrac{Z^2}{A^{4/3}}.
\end{multline}

The components of  $K(\rho_0(x))$ in Eq.~(\ref{eq:15}) are as follows:
\begin{itemize}
  \item [(i)]
  \begin{equation}\label{eq:16}
   K_{NM}(\rho_0(x))=9\rho^2_0(x)\left.\dfrac{\partial^2  \Big(\frac{\varepsilon(\rho,\delta=0)}{\rho}\Big)}{\partial\rho^2}\right|_{\rho=\rho_0(x)}
  \end{equation}
  is the incompressibility of \emph{symmetric} NM (known also as $K_{\infty}$ or ``the volume term'' $K_{vol}$, see Ref.~\cite{Colo2014});
  \item [(ii)] The asymmetry term of the NM incompressibility:
  \begin{align}\label{eq:17}
  K_{\tau}(\rho_0(x))=&K_{sym}(\rho_0(x))-6L_{sym}(\rho_0(x))\notag\\ &-\dfrac{L_{sym}(\rho_0(x))Q(\rho_0(x))}{K_{NM}(\rho_0(x))}
  \end{align}
  or
  \begin{align}\label{eq:18}
  K_{\tau}(\rho_0(x))=&K_{sym}(\rho_0(x))+3L_{sym}(\rho_0(x))\notag\\&-L_{sym}(\rho_0(x))B(\rho_0(x)),
  \end{align}
  where
  \begin{gather}\label{eq:19}
    K_{sym}(\rho_0(x))=9\rho^2_0(x)\left.\dfrac{\partial^2 S(\rho)}{\partial\rho^2}\right|_{\rho=\rho_0(x)}.
  \end{gather}
  The function $L_{sym}$ in Eqs.~(\ref{eq:17}) and~(\ref{eq:18}) is known as a ``slope parameter'' and has the form:
  \begin{gather}\label{eq:20}
  L_{sym}(\rho_0(x))=3\rho_0(x)\left.\dfrac{\partial S(\rho)}{\partial\rho}\right|_{\rho=\rho_0(x)}.
  \end{gather}
  The functions $Q(\rho_0(x))$ and $B(\rho_0(x))$ in Eqs.~(\ref{eq:17}) and~(\ref{eq:18}) are defined as follows:
  \begin{equation}\label{eq:21}
  Q(\rho_0(x))=27\rho^3_0(x)\left.\dfrac{\partial^3 \Big(\frac{\varepsilon(\rho,\delta=0)}{\rho}\Big) }{\partial\rho^3}\right|_{\rho=\rho_0(x)}
  \end{equation}
  and
  \begin{equation}\label{eq:22}
  B(\rho_0(x))=\dfrac{27\rho_0^2}{K_{NM}(\rho_0(x))}\left.\dfrac{\partial^3{\varepsilon}(\rho,\delta=0)}{\partial\rho^3}\right|_{\rho=\rho_0(x)}\!\!\!\!\!\!\!\!\!\!\!\!\!\!.
  \end{equation}
  The latter function can be written also as:
  \begin{equation}\label{eq:23}
  B(\rho_0(x))=9+\dfrac{Q(\rho_0(x))}{K_{NM}(\rho_0(x))}.
  \end{equation}
  In our previous work~\cite{Gaidarov2023} we have used the notation $\Delta K^{NM}$ instead of $K_{sym}$ from Eq.~(\ref{eq:19}).
  \item [(iii)] The Coulomb term is given by:
  \begin{equation}\label{eq:24}
  K_{Coul}(\rho_0(x))=\dfrac{3}{5}\dfrac{e^2}{r_0(x)}(1-B(\rho_0(x)))
  \end{equation}
  with $e^2\cong1.44$ MeV.fm
  and
  \begin{equation}\label{eq:26}
  r_0(x)=\Big(\dfrac{3}{4\pi\rho_0(x)}\Big)^{1/3}.
  \end{equation}
\end{itemize}

Here we should mention the formal partial similarity of
Eq.~(\ref{eq:15}) with Eq.~(\ref{eq:17}) in Ref.~\cite{Colo2014}.
The latter gives the  incompressibility for finite nuclei, while
our Eq.~(\ref{eq:15}) is for the incompressibility in nuclear
matter. Eq.~(\ref{eq:17}) in Ref.~\cite{Colo2014} accounts for the
surface contribution, while in our case Eq.~(\ref{eq:14}) makes a
transition from $K(\rho_0(x))$ in \emph{NM} to $K^A$ in
\emph{finite nuclei}, accounting for the surface contributions
weighting $K(\rho_0(x))$ by the weight function $|F(x)|^2$, which
is related to the density distribution of a given nucleus.

\subsection{Brueckner and Barcelona-Catania-Paris-Madrid EDFs}
\label{subsec:EDF}

In what follows we give the expressions for the EDFs used in the
present work. We consider the energy as a sum of the kinetic
$T(\rho,\alpha)$ and the potential $V(\rho,\alpha)$ contributions
\begin{equation}\label{eq:27}
  \frac{E(\rho,\alpha)}{A}=T(\rho,\alpha)+V(\rho,\alpha),\quad \alpha=\frac{N-Z}{A}
\end{equation}
in both approaches for EDF’s, namely that one of
Brueckner~\cite{Brueckner68, Brueckner69, Brueckner68a} and of the
Barcelona-Catania-Paris-Madrid (BCPM) one (see, e.g.
Ref.~\cite{Baldo2023,Baldo2013,Baldo2017} and references therein).

The kinetic energy part of the EDF used is of the Thomas--Fermi
type:
\begin{equation}\label{eq:28}
T(\rho,\alpha)=\frac{C}{2}[(1+\alpha)^{5/3}+(1-\alpha)^{5/3}]\rho^{2/3}.
\end{equation}
We use two types of the potential part of the EDF $V(\rho,\alpha)$:
\begin{itemize}
  \item [i)] That one from the Brueckner EDF~\cite{Brueckner69, Brueckner68a}:
  \begin{align}\label{eq:29}
  V(\rho,\alpha)=&b_1\rho+b_2\rho^{4/3}+b_3\rho^{5/3}\notag\\&+\alpha^2(b_4\rho + b_5 \rho^{4/3} + b_6 \rho^{5/3})
  \end{align}
and
  \item [ii)] The potential part of the BCPM EDF (see, e.g. Ref.~\cite{Baldo2023}):
  \begin{align}\label{eq:30}
  V(\rho,\alpha)=\sum_{n=1}^{5} a_n\Big(\frac{\rho}{\rho_{\infty}}\Big)^n (1-\alpha^2) + \sum_{n=1}^{5} b_n\Big(\frac{\rho}{\rho_{0n}}\Big)^n \alpha^2.
  \end{align}
\end{itemize}
The values of the parameters $a_n$ and $b_n$ in Eq.~(\ref{eq:30}),
as well as $\rho_{\infty} = 0.16$~fm$^{-3}$ and $\rho_{0n} =
0.155$~fm$^{-3}$ are given in Ref.~\cite{Baldo2023}.

In Eqs.~(\ref{eq:28}) and (\ref{eq:29}) the values of the
parameters $C$ and $b_1\div b_6$ are listed in Appendix~\ref{appA}
[see Eqs.~(\ref{eq:A6}) and (\ref{eq:A7})]. The expressions for
the quantities $S(\rho_0(x))$ [Eq.~(\ref{eq:12})],
$K_{NM}(\rho_0(x))$ [Eq.~(\ref{eq:16})], $K_{sym}(\rho_0(x))$
[Eq.~(\ref{eq:19})], $L_{sym}(\rho_0(x))$ [Eq.~(\ref{eq:20})],
$Q(\rho_0(x))$ [Eq.~(\ref{eq:21})] and $B(\rho_0(x))$
[Eqs.~(\ref{eq:22}) and (\ref{eq:23})] obtained in our work for
the cases of the Brueckner \cite{Brueckner68, Brueckner69,
Brueckner68a} and BCPM EDF (see, e.g. Ref.~\cite{Baldo2023} and
references therein), are given in Appendix~\ref{appA}.

\section{Results and discussion}\label{sec:results}

In the calculations the density distributions $\rho(r)$, which are
necessary to compute the weight function $|F(x)|^{2}$
[Eq.~(\ref{eq:9})], are obtained within the self-consistent
deformed Hartree-Fock method with density-dependent SLy4
interaction \cite{sly4} and pairing correlations~\cite{vautherin}
(see also Refs.~\cite{Gaidarov2011, Gaidarov2012, Gaidarov2014,
Antonov2016, Danchev2020, Sarriguren2007}). The expressions for
the single-particle wave functions and densities $\rho_{p,n}(r)$
in the mentioned method are given, e.g., in
Ref.~\cite{Gaidarov2011}.

The mean square radii for protons and neutrons are obtained using
the corresponding definitions:
\begin{equation}
<r_{p,n}^2> =\frac{ \int r^2\rho_{p,n}({\bf r})d{\bf r}} {\int
\rho_{p,n}({\bf r})d{\bf r}},
\label{eq:32}
\end{equation}
while the matter mean square radius $<r^{2}>$ entering
Eq.~(\ref{eq:2}) can be calculated by the expression:
\begin{equation}
<r^{2}>=\frac{N}{A}<r_{n}^{2}>+\frac{Z}{A}<r_{p}^{2}>.
\label{eq:33}
\end{equation}

We start our analysis by calculating the NSE and some parameters
related to its density dependence. The values of the symmetry
energy $S^{A}$, the slope parameter $L_{sym}^{A}$, and skewness
parameter $Q^{A}$ in a wide range of nuclei from $^{40}$Ca to
$^{208}$Pb obtained using EDF's of Brueckner and $V(\rho,\alpha)$
from BCPM [further called BCPM(v)] are given in
Table~\ref{tab:table1}. Similarly to Eqs.~(\ref{eq:13}) and
(\ref{eq:14}), the skewness parameter $Q^A$ for finite nuclei can
be calculated in the CDFM scheme. The values of the symmetry
energies $S^{A}$ deduced from the calculations with both EDF's are
realistic lying in the interval between 25 and 30 MeV. Concerning
the values of the slope parameter, it can be seen from
Table~\ref{tab:table1} that the BCPM(v) functional produces almost
twice larger values compared with the $L^A_{sym}$ values obtained
using the Brueckner EDF. For instance, the calculations based on
interactions derived from chiral effective field theory (EFT)
predict for $^{208}$Pb nucleus values for the symmetry energy of
$31.3\pm0.8$ MeV and slope parameter of $52.6\pm4.0$ MeV
\cite{Sammarruca2022}. The use of the BCPM(v) functional leads to
values for these two characteristics of $^{208}$Pb, which are very
close to the ones obtained in Ref.~\cite{Sammarruca2022} (see
Table~\ref{tab:table1}). Therefore, this is a good starting point
to study further quantities such as the intrinsic symmetry energy
parameter, namely the incompressibility $K^{A}$, and the mean
square nuclear radius $<r^{2}>$ that are basic ingredients to
calculate the monopole excitation energy.

\begin{table}[h]
\caption{\label{tab:table1} The values of the symmetry energy
$S^{A}$ [Eq.~(\ref{eq:13})], slope parameter $L^A_{sym}$
[Eq.~(\ref{eq:20})], and skewness parameter $Q^A$ of the
considered nuclei calculated within the CDFM by using Brueckner
and BCPM(v) energy-density functionals.}
\begin{ruledtabular}
\begin{tabular}{ccccccc}
&\multicolumn{3}{c}{Brueckner}&\multicolumn{3}{c}{BCPM(v)}\\
\cline{2-4}\cline{5-7}\\[-8pt]
Nuclei&$S^{A}$&$L^A_{sym}$&$Q^A$&$S^{A}$&$L^A_{sym}$&$Q^A$\\
      &[MeV]  &[MeV]      &[MeV]      &[MeV]  &[MeV]      &[MeV]\\
\hline\\[-8pt]
 $^{40}$Ca&25.61&25.16&-373.66&27.87&53.19&-257.59\\
 $^{42}$Ca&25.86&25.13&-379.65&28.17&53.65&-264.48\\
 $^{44}$Ca&26.11&25.01&-385.80&28.45&54.10&-271.41\\
 $^{46}$Ca&26.35&25.22&-390.29&28.66&54.36&-289.03\\
 $^{48}$Ca&26.66&24.92&-399.06&29.03&54.97&-298.20\\
 $^{54}$Fe&27.34&24.82&-415.00&29.78&56.13&-322.20\\
 $^{56}$Ni&27.58&24.64&-421.61&30.06&56.58&-330.95\\
 $^{58}$Ni&27.24&24.29&-416.26&29.73&56.08&-315.65\\
 $^{60}$Ni&26.98&24.02&-412.14&29.47&55.67&-305.05\\
 $^{68}$Ni&27.00&25.67&-403.01&29.28&55.07&-326.62\\
 $^{64}$Zn&26.65&24.57&-401.57&29.03&54.85&-300.56\\
 $^{68}$Zn&26.75&25.40&-399.03&29.05&54.75&-311.96\\
 $^{90}$Zr&27.94&28.56&-408.56&29.85&55.21&-419.70\\
 $^{92}$Zr&27.68&27.94&-406.37&29.67&55.07&-397.63\\
 $^{92}$Mo&28.02&28.54&-410.60&29.95&55.36&-422.09\\
 $^{94}$Mo&27.79&28.20&-407.30&29.74&55.10&-407.35\\
 $^{96}$Mo&27.59&27.93&-404.35&29.55&54.85&-395.98\\
$^{106}$Cd&27.93&28.73&-407.78&29.81&55.06&-423.52\\
$^{110}$Cd&27.90&29.12&-404.73&29.72&54.86&-426.21\\
$^{112}$Cd&27.89&29.18&-404.20&29.72&54.84&-425.20\\
$^{114}$Cd&27.89&29.41&-402.87&29.68&54.73&-429.10\\
$^{116}$Cd&27.88&29.58&-401.85&29.64&54.64&-432.47\\
$^{100}$Sn&28.59&29.28&-419.35&30.44&55.94&-461.83\\
$^{112}$Sn&28.12&29.25&-409.10&29.95&55.17&-438.16\\
$^{114}$Sn&28.12&29.51&-407.63&29.91&55.05&-442.58\\
$^{116}$Sn&28.12&29.70&-406.48&29.87&54.95&-446.28\\
$^{118}$Sn&28.12&29.65&-406.71&29.88&54.98&-444.29\\
$^{120}$Sn&28.12&29.78&-406.20&29.87&54.92&-447.63\\
$^{122}$Sn&28.15&29.75&-407.02&29.91&54.98&-447.61\\
$^{124}$Sn&28.21&29.96&-407.10&29.93&54.96&-454.81\\
$^{132}$Sn&28.63&30.70&-412.58&30.26&55.26&-490.43\\
$^{204}$Pb&28.93&33.65&-403.13&30.17&54.38&-544.21\\
$^{206}$Pb&28.93&33.44&-404.28&30.20&54.45&-541.14\\
$^{208}$Pb&28.91&33.27&-404.90&30.20&54.47&-539.23
\end{tabular}
\end{ruledtabular}
\end{table}

In Table~\ref{tab:table2} are shown the values of the
incompressibility $K^{A}$ and its components $K^{A}_{NM}$,
$K^{A}_{\tau}$ and $K^{A}_{Coul}$ in the same range of nuclei
calculated using EDF of Brueckner and BCMP(v). While the
differences of the symmetry energy $S^{A}$ for both EDFs which can
be seen in Table~\ref{tab:table1} are relatively small (2.26 MeV,
2.49 MeV, 1.88 MeV for $^{40}$Ca, $^{56}$Ni, and $^{106}$Cd,
respectively) there are large differences in the value of $K^{A}$.

\begin{table*}
\caption{\label{tab:table2}The values of nuclear properties,
including $K^A_{NM}$, $K^A_{\tau}$, $K^{A}_{Coul}$, $K^{A}$, and
$E_{ISGMR}$ calculated using Brueckner and BCPM(v) EDFs and
Eq.~(\ref{eq:2}), along with rms radii and experimental data for
the ISGMR energy.}
\begin{ruledtabular}
\begin{tabular}{ccccccccccccc}
&\multicolumn{5}{c}{Brueckner}&\multicolumn{5}{c}{BCPM(v)}&&\\
\cline{2-6}\cline{7-11}\\[-8pt]
Nuclei&$K^A_{NM}$&$K^A_{\tau}$&$K^{A}_{Coul}$&$K^{A}$&$E_{ISGMR}$&$K^A_{NM}$&$K^A_{\tau}$&$K^{A}_{Coul}$&$K^{A}$&$E_{ISGMR}$&$\langle r^2\rangle^{1/2}$&Exp.\\
      &[MeV]  &[MeV] &[MeV] &[MeV] &[MeV] &[MeV]  &[MeV] &[MeV] &[MeV] &[MeV] &[fm] &[MeV]\\
\hline\\[-8pt]
 $^{40}$Ca&118.78&-367.82&-3.50&108.53&19.76&184.27&-280.05&-4.01&172.53&24.92&3.40&$19.18\pm0.37$ \cite{Anders2013}\\
 $^{42}$Ca&120.46&-371.88&-3.52&109.98&19.67&187.10&-280.80&-4.01&175.47&24.85&3.43&$19.7\pm0.1$   \cite{Howard2019}\\
 $^{44}$Ca&122.15&-375.88&-3.54&109.94&19.45&189.81&-281.43&-4.02&177.14&24.69&3.47&$19.49\pm0.34$ \cite{Button2017}\\
 $^{46}$Ca&123.58&-377.80&-3.55&108.54&19.12&190.49&-279.04&-4.01&176.02&24.35&3.51&--\\
 $^{48}$Ca&125.90&-383.17&-3.57&107.08&18.81&193.97&-279.73&-4.01&176.99&24.18&3.54&$19.88\pm0.16$ \cite{Anders2013}\\
 $^{54}$Fe&130.40&-393.27&-3.61&117.90&19.20&200.53&-280.36&-4.02&186.83&24.17&3.64&$19.66\pm0.37$ \cite{Button2019}\\
 $^{56}$Ni&132.18&-397.24&-3.63&118.91&19.12&202.98&-280.54&-4.03&188.24&24.05&3.67&$19.1\pm0.5$ \cite{Bagchi2015}\\
 $^{58}$Ni&130.44&-393.37&-3.59&117.42&18.78&201.13&-280.61&-4.01&186.78&23.69&3.72&$18.43\pm0.15$ \cite{Lui2006}\\
 $^{60}$Ni&129.10&-390.27&-3.57&115.47&18.42&199.63&-280.31&-4.00&185.05&23.32&3.76&$17.62\pm0.15$ \cite{Lui2006}\\
 $^{68}$Ni&127.54&-385.47&-3.58&105.43&16.80&196.32&-275.46&-3.95&176.58&21.74&3.94&$21.1\pm1.9$ \cite{Vandebrouck2014,Vandebrouck2015}\\
 $^{64}$Zn&126.45&-383.49&-3.54&112.50&17.75&195.54&-277.91&-3.96&180.52&22.48&3.85&$18.88\pm0.79$ \cite{Button2019}\\
 $^{68}$Zn&126.23&-382.87&-3.55&109.41&17.13&195.23&-276.44&-3.95&178.61&21.88&3.93&$16.6\pm0.17$ \cite{Button2019}\\
 $^{90}$Zr&131.04&-387.67&-3.64&111.82&15.99&196.71&-260.62&-3.83&178.32&20.19&4.26&$16.9\pm0.1$ \cite{Howard2019}\\
 $^{92}$Zr&130.00&-386.61&-3.62&109.49&15.69&196.75&-263.71&-3.84&177.47&19.97&4.30&$16.5\pm0.1$ \cite{Howard2019}\\
 $^{92}$Mo&131.60&-389.18&-3.64&113.18&15.99&197.88&-260.94&-3.82&179.66&20.15&4.29&$16.6\pm0.1$ \cite{Howard2019}\\
 $^{94}$Mo&130.44&-386.94&-3.62&111.11&15.71&196.79&-262.17&-3.83&178.02&19.89&4.32&$16.4\pm0.2$ \cite{Howard2019}\\
 $^{96}$Mo&129.44&-384.84&-3.61&108.95&15.44&195.70&-262.90&-3.83&176.23&19.63&4.36&$16.3\pm0.2$ \cite{Howard2019}\\
$^{106}$Cd&130.94&-386.97&-3.63&110.82&15.12&196.84&-259.46&-3.80&177.07&19.11&4.48&$16.27\pm0.09$ \cite{Patel2012}\\
$^{110}$Cd&130.31&-385.31&-3.63&108.20&14.74&195.84&-258.52&-3.79&175.09&18.75&4.55&$15.94\pm0.07$ \cite{Patel2012}\\
$^{112}$Cd&130.20&-385.32&-3.63&106.83&14.55&196.02&-258.86&-3.79&174.57&18.60&4.58&$15.80\pm0.05$ \cite{Patel2012}\\
$^{114}$Cd&129.96&-384.33&-3.63&105.24&14.35&195.24&-257.80&-3.78&173.06&18.40&4.61&$15.61\pm0.08$ \cite{Patel2012}\\
$^{116}$Cd&129.78&-383.52&-3.63&103.59&14.15&194.60&-256.86&-3.77&171.60&18.21&4.63&$15.44\pm0.06$ \cite{Patel2012}\\
$^{100}$Sn&134.59&-394.20&-3.69&114.73&15.72&200.14&-256.55&-3.81&179.65&19.67&4.39&--\\
$^{112}$Sn&131.64&-388.16&-3.65&110.29&14.82&197.56&-257.83&-3.79&177.06&18.78&4.56&$16.2\pm0.1$ \cite{Cao2012}\\
$^{114}$Sn&131.39&-387.09&-3.65&109.05&14.65&196.74&-256.66&-3.78&175.78&18.60&4.59&$16.1\pm0.1$ \cite{Cao2012}\\
$^{116}$Sn&131.20&-386.23&-3.65&107.73&14.47&196.06&-255.66&-3.77&174.54&18.42&4.62&$15.8\pm0.1$ \cite{Cao2012}\\
$^{118}$Sn&131.22&-386.61&-3.65&106.47&14.30&196.51&-256.22&-3.77&174.26&18.30&4.65&$15.8\pm0.1$ \cite{Cao2012}\\
$^{120}$Sn&131.16&-386.10&-3.65&105.03&14.13&196.07&-255.41&-3.76&173.07&18.13&4.67&$15.7\pm0.1$ \cite{Cao2012}\\
$^{122}$Sn&131.38&-386.80&-3.65&103.71&13.96&196.68&-255.76&-3.77&172.81&18.02&4.70&$15.4\pm0.1$ \cite{Cao2012}\\
$^{124}$Sn&131.54&-386.61&-3.65&102.28&13.80&196.36&-254.40&-3.76&171.64&17.87&4.72&$15.3\pm0.1$ \cite{Cao2012}\\
$^{132}$Sn&133.59&-389.61&-3.69& 96.99&13.19&197.57&-249.47&-3.73&169.05&17.42&4.81&--\\
$^{204}$Pb&132.82&-384.32&-3.70& 97.31&11.50&195.57&-239.26&-3.59&166.26&15.03&5.53&$13.98$ \cite{Fujiwara2011}\\
$^{206}$Pb&133.01&-385.04&-3.70& 96.55&11.42&196.16&-240.01&-3.60&166.31&14.99&5.54&$13.94$ \cite{Fujiwara2011}\\
$^{208}$Pb&133.08&-385.18&-3.70& 95.67&11.34&196.23&-240.22&-3.60&165.85&14.93&5.56&$13.96\pm0.2$ \cite{Youngblood2004}
\end{tabular}
\end{ruledtabular}
\end{table*}

The corresponding energy $E_{ISGMR}$ calculated in the present
work using Eq.~(\ref{eq:2}) are also given in
Table~\ref{tab:table2} and are compared with the available
experimental data for the same range of nuclei. We should mention
that the results for the centroid energy in the case of the
Brueckner EDF are in good agreement with the data for isotopes of
Ca, Fe, and Ni and acceptable for those of Zn, Mo, and Cd. In the
case of BCMP(v) EDF they are comparable in the case of $^{68}$Ni
and partly for some isotopes of Pb.  The use of Eq.~(\ref{eq:2})
includes the calculations of two quantities, namely, the
incompressibility $K^{A}(N,Z)$ which is obtained  by
Eqs.~(\ref{eq:14})-(\ref{eq:19}) and (\ref{eq:24})-(\ref{eq:26}),
and the rms radii $<r^{2}>^{1/2}$ which are calculated by
Eqs.~(\ref{eq:32}) and (\ref{eq:33}) using density distributions
obtained within the self-consistent deformed Hartree-Fock method
with SLy4 effective Skyrme force and with pairing correlations.

Concerning the use of Eq.~(\ref{eq:2}), in particular the nuclear
rms radii in it, we would like to note the following: i) the rms
radii can be calculated on the base of experimentally obtained
proton density distributions by means of electromagnetic
interaction and some other processes; ii) while in many cases the
charge densities are well known (see, e.g., Ref.~\cite{DeVries87})
with relatively high accuracy, this is not the case with the
neutron densities which could not be precisely obtained due to the
charge neutrality of the neutrons. Thus, there is generally a lack
of data for the neutron density, or their knowledge comes only
from a small number of specific experiments. In the majority of
cases the neutron density is used on the base of some
approximations of the proton densities. In many cases (as in the
present work) estimations of the proton and neutron densities
obtained in various theoretical methods are used.

The above mentioned points (i) and (ii) are particularly related
to the neutron-rich nuclei, which are the majority of the
considered nuclei. Among the latter there are four neutron-rich
exotic nuclei with short half-life time, namely $^{56}$Ni,
$^{68}$Ni, $^{100}$Sn, and $^{132}$Sn. The reasons mentioned above
lead to certain difficulties, namely reliable total nucleon
density distributions ($\rho=\rho_{p}+\rho_{n}$) to be used in the
calculations of the rms radii which enter Eq.~(\ref{eq:2}).

\begin{figure}[h]
\centering
\includegraphics[width=86mm]{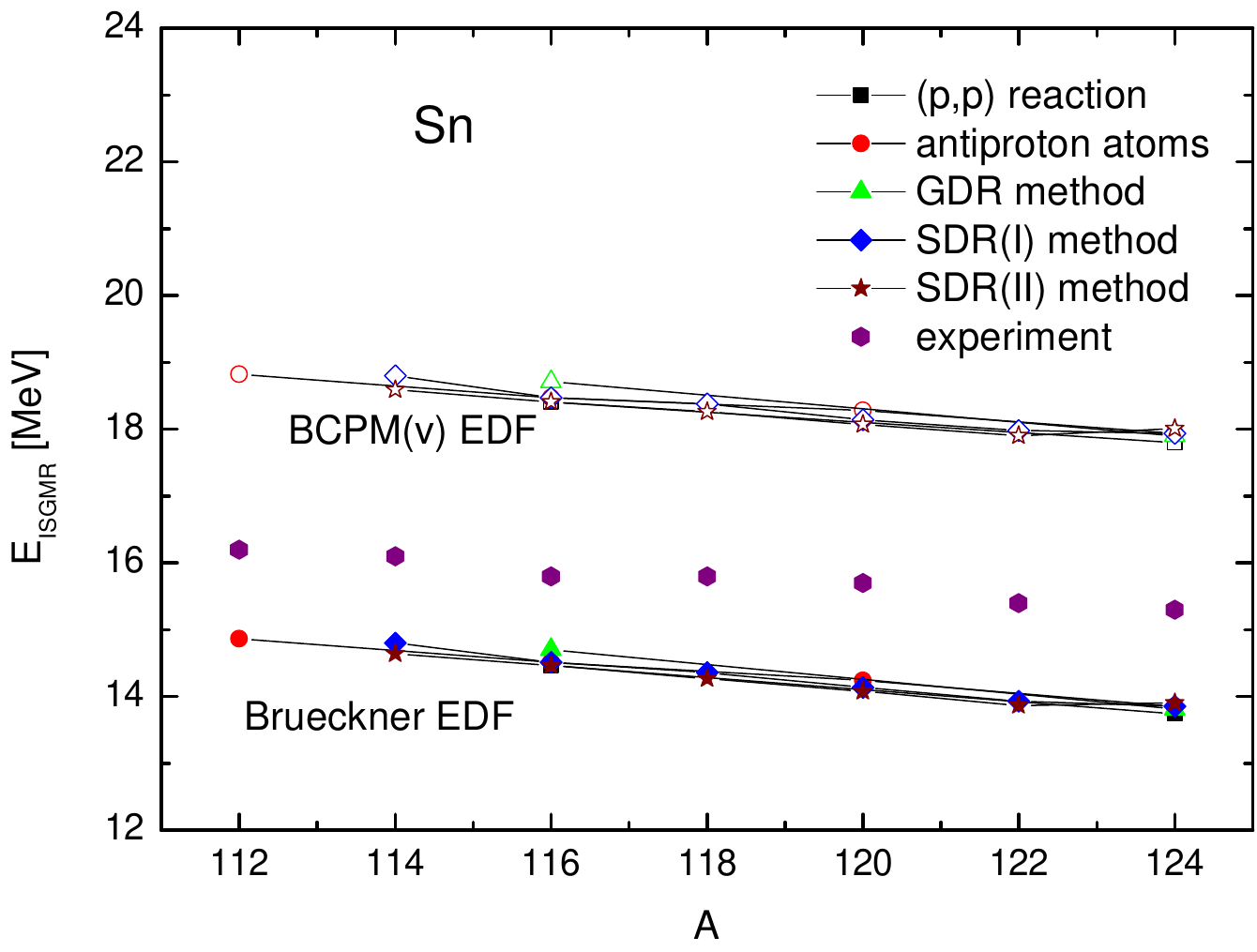}
\caption[]{The centroid energies $E_{ISGMR}$ as a function of the
mass number A of Sn isotopes ($A=112-124$) obtained with Brueckner
(full symbols)  and BCPM(v) (open symbols) EDFs by using the
information for the neutron skin thickness extracted from various
experiments including $(p,p)$ scattering (squares), antiprotonic
atoms (circle), the giant dipole resonance (GDR) method
(triangles), and the spin-dipole resonance (SDR) method (diamonds
and stars). The experimental data (hexagons) are taken from
Ref.~\cite{Cao2012}. \label{fig1}}
\end{figure}

Along this line our next step is to consider the rare cases, where
data extracted for the neutron skin thickness with probes having
different sensitivities to the proton and neutron distributions
are available. As an example, we analyze the Sn isotopic chain
($A=112-124$). Usually the neutron skin thickness in heavy nuclei
is defined as
\begin{equation}
\Delta r_{np}=<r_{n}^{2}>^{1/2}-<r_{p}^{2}>^{1/2}.
\label{eq:34}
\end{equation}
The "semi-empirical" procedure aims to determine the mean square
mass radius $<r^{2}>$ entering Eq.~(\ref{eq:2}) starting from the
experimental values of $\Delta r_{np}$. They are measured in
different methods
\cite{Ray79,Hoffmann81,Trzcinska2001,Krasz94,Krasz99,Krasz2004}
and shown in Fig.~4 of Ref.~\cite{Sarriguren2007}. As known, a
measurement of the neutron density distributions to a precision
and detail comparable to that of the proton one is hardly
possible. Having the proton rms radii calculated from the charge
radii determined with a high accuracy \cite{DeVries87,
Patterson2003}, the neutron rms radii can be obtained from
Eq.~(\ref{eq:34}). Then, the matter mean square radius $<r^{2}>$
are calculated using Eq.~(\ref{eq:33}). The values of the rms
radius $<r^{2}>^{1/2}$ deduced from the procedure are listed in
Table~\ref{tab:table3}. It is seen that the values of
$<r^{2}>^{1/2}$ are very close to those for the same Sn isotopes
obtained within the HF+BCS calculations with SLy4 interaction and
given in Table~\ref{tab:table2}. The resulting centroid energies
$E_{ISGMR}$ calculated using Eq.~(\ref{eq:2}) are displayed in
Fig.~\ref{fig1}. In the same figure a comparison with the
available experimental data \cite{Cao2012} is made. Both Brueckner
and BCPM(v) functionals provide excitation energies whose values
almost coincide with the calculated ones given in
Table~\ref{tab:table2}. The small differences in the excitation
energy values represent a clear indication of the minimal role of
the radius calculated from the empirical data for the neutron
skins in Sn isotopes. However, we would like to note the
importance of such kind of "alternative" estimations (where
possible) avoiding the ambiguities in determination of the neutron
rms radii both experimentally and theoretically.

\begin{table}[h]
\caption{\label{tab:table3} The values of the rms radius
$<r^{2}>^{1/2}$ [Eq.~(\ref{eq:33})] of Sn isotopes ($A=112-124$)
calculated in the procedure based on measurements of the neutron
skin thickness $\Delta r_{np}$ [Eq.~(\ref{eq:34})] of the same
nuclei.}
\begin{ruledtabular}
\begin{tabular}{ccc}
Exp. method & Nuclei & $<r^{2}>^{1/2}$\\[1pt]
            &        &  [fm]\\
\hline\\[-8pt]
$(p,p)$ reaction \cite{Ray79,Hoffmann81}              &  $^{116}$Sn  &  $4.624\pm0.029$\\
                                                      &  $^{124}$Sn  &  $4.741\pm0.030$\\
antiproton atoms \cite{Trzcinska2001}                 &  $^{112}$Sn  &  $4.554\pm0.011$\\
                                                      &  $^{116}$Sn  &  $4.607\pm0.012$\\
                                                      &  $^{120}$Sn  &  $4.635\pm0.012$\\
                                                      &  $^{124}$Sn  &  $4.704\pm0.012$\\
giant dipole resonance \cite{Krasz94}                 &  $^{116}$Sn  &  $4.549\pm0.068$\\
                                                      &  $^{124}$Sn  &  $4.716\pm0.067$\\
spin dipole resonance (I) \cite{Krasz99,Krasz2004}    &  $^{114}$Sn  &  $4.544\pm0.028$\\
                                                      &  $^{116}$Sn  &  $4.607\pm0.035$\\
                                                      &  $^{118}$Sn  &  $4.628\pm0.035$\\
                                                      &  $^{120}$Sn  &  $4.671\pm0.041$\\
                                                      &  $^{122}$Sn  &  $4.709\pm0.042$\\
                                                      &  $^{124}$Sn  &  $4.704\pm0.042$\\
spin dipole resonance (II) \cite{Krasz99,Krasz2004}   &  $^{114}$Sn  &  $4.594\pm0.006$\\
                                                      &  $^{116}$Sn  &  $4.624\pm0.006$\\
                                                      &  $^{118}$Sn  &  $4.658\pm0.005$\\
                                                      &  $^{120}$Sn  &  $4.689\pm0.006$\\
                                                      &  $^{122}$Sn  &  $4.780\pm0.006$\\
                                                      &  $^{124}$Sn  &  $4.686\pm0.006$\\
\end{tabular}
\end{ruledtabular}
\end{table}

As additional examples of this procedure, we consider two more
nuclei, $^{208}$Pb and $^{48}$Ca, for which the neutron skin
thickness has been recently extracted from parity-violating
experiments PREX-II \cite{Adhikari2021} and CREX \cite{CREX2022},
correspondingly. The relatively large neutron skin thickness
$\Delta r_{np}$($^{208}$Pb)=$0.283\pm 0.071$ fm extracted from
PREX-II and the small value $\Delta r_{np}$($^{48}$Ca)=0.121$\pm
0.026$$\pm 0.024$ fm extracted from CREX are discussed in
Ref.~\cite{Machleidt2024}. In this sense, we want to see the
influence of the differences between PREX-II and CREX on the
monopole excitations in $^{208}$Pb and $^{48}$Ca nuclei. The
values of the rms radii deduced from the procedure are
$<~r^{2}>^{1/2}$($^{48}$Ca)=3.432 fm and
$<r^{2}>^{1/2}$($^{208}$Pb)=5.608 fm, respectively, which are
close to the calculated values listed in Table~\ref{tab:table2}.
For $^{48}$Ca nucleus this leads to centroid energies $E_{ISGMR}$
of $19.42\pm0.170$ MeV and $24.97\pm0.170$ MeV when applying
Brueckner and BCPM(v) EDF's, while for $^{208}$Pb the
corresponding values are $11.24\pm0.088$ MeV and $14.80\pm0.116$
MeV. It can be seen that, taking into account the estimated error
in the case of Brueckner EDF, the extracted value of the
$E_{ISGMR}$ for $^{48}$Ca, using the neutron skin thickness
reported by the CREX collaboration, fits well with the
experimental data.

At this point we turn to the general question, namely, which one,
Eq.~(\ref{eq:1}) or Eq.~(\ref{eq:2}) to be used in the
calculations of $E_{ISGMR}$. The point is that there are two
different quantities related to the size of the nucleus in
Eq.~(\ref{eq:1}) and Eq.~(\ref{eq:2}), namely, $r_{0}A^{1/3}$ in
Eq.~(\ref{eq:1}) and the root-mean-square (rms) radius
$<r^{2}>^{1/2}$ in Eq.~(\ref{eq:2}). For this purpose we give in
what follows the results for $E_{ISGMR}$ obtained from
calculations using Eq.~(\ref{eq:1}), where the nuclear radius is
given by $R=r_{0} A^{1/3}$, $r_{0}$ being a radial parameter of
density distributions obtained from different experiments. Using
the comparison of $E_{ISGMR}$ with the experimental data, we
performed a more detailed semi-empirical analysis of the parameter
$r_{0}$. We parameterized it in a form:
\begin{equation}
r_{0}=(1 + x/A^{y})
\label{eq:35}
\end{equation}
and obtained values of $x$ and $y$ from the fit of $E_{ISGMR}$
with the data for two nuclei in both limits of the considered
nuclear range, namely for $^{40}$Ca and $^{208}$Pb. In the case of
$^{40}$Ca $r_{0}$=1.205 fm and in $^{208}$Pb $r_{0}$=1.068 fm.
These two conditions lead to the approximated values of $x$=2.40
fm and $y$=2/3. Thus, we obtained the following expression for the
radius in the denominator of Eq.~(\ref{eq:1}):
\begin{equation}
R=r_{0}A^{1/3}=[1+2.40/A^{2/3}]A^{1/3}.
\label{eq:36}
\end{equation}
This equation can be used for the same range of nuclei. In our
opinion, the second term in the right-hand side of
Eq.~(\ref{eq:36}) $2.40/A^{1/3}$ is related to the diffuse nuclear
surface which changes its range from medium to heavy nuclei. Let
us remind at this point the behavior of the nuclear energy per
particle $B/A$ $(B/A=a_{1}+a_{2}/A^{1/3}+...$, with the known
parameters $a_{1}$ and $a_{2}$) with its volume and surface terms
in the right-hand side of this expression. Here we would like to
note that our expressions (\ref{eq:35}) and (\ref{eq:36}) are for
the parameter $r_{0}$, while in Ref.~\cite{Angeli2013} is given an
expansion for the charge rms radius.

\begin{figure}[h]
\centering
\includegraphics[width=86mm]{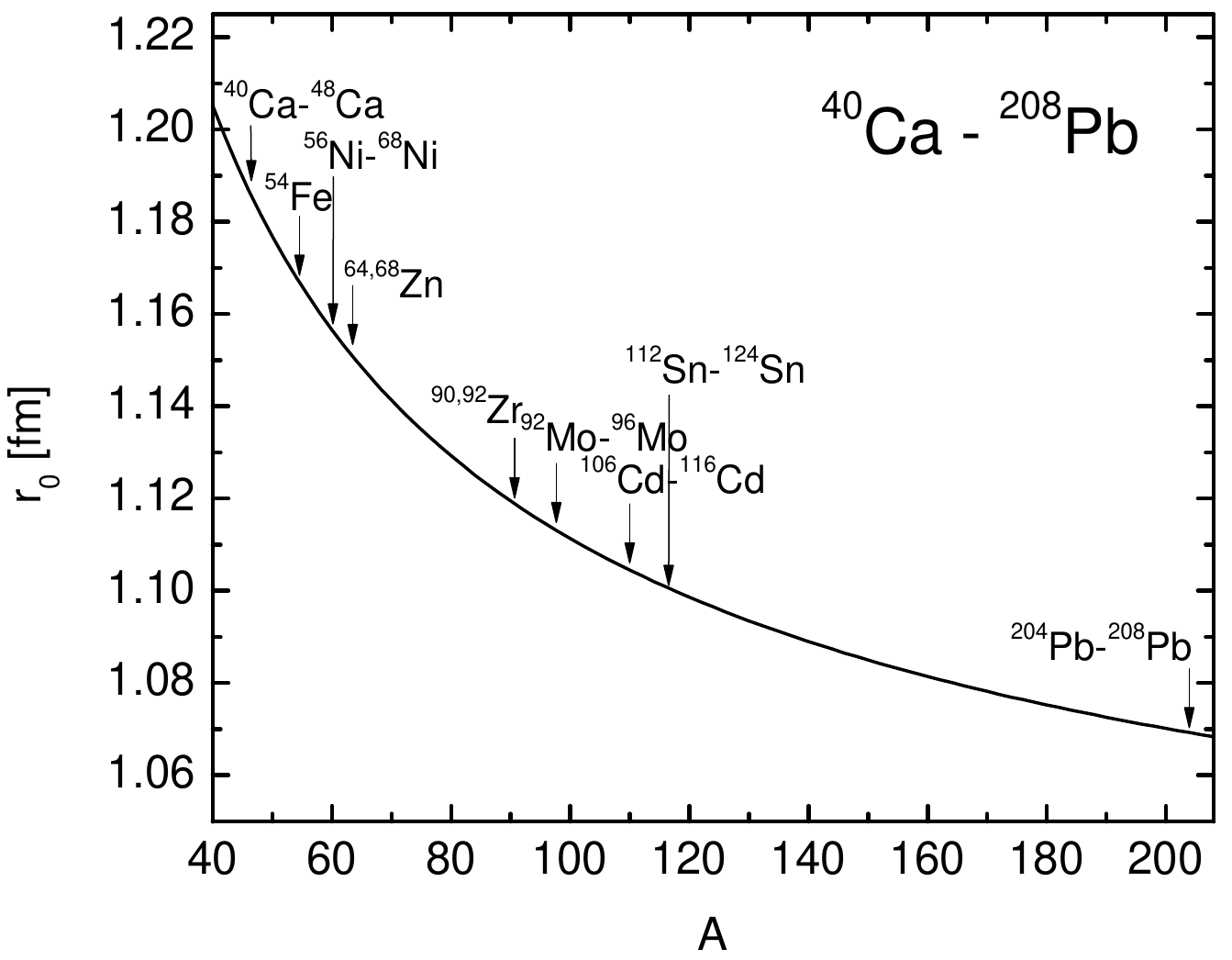}
\caption[]{The dependence of the radius parameter $r_{0}$ (in fm)
entering Eq.~(\ref{eq:1}) on the mass number A for the considered
nuclei between $^{40}$Ca and $^{208}$Pb.
\label{fig2}}
\end{figure}

The dependence of the radial parameter $r_{0}=1+2.40/A^{2/3}$ on
the mass number between $^{40}$Ca and $^{208}$Pb is shown in
Fig.~\ref{fig2}. The results of the calculations of $E_{ISGMR}$
using Eqs.~(\ref{eq:1}), (\ref{eq:35}) and (\ref{eq:36}) in both
cases [with the EDFs of Brueckner and BCMP(v)] are listed in
Table~\ref{tab:table4}. It can be seen that in this case the
results of $E_{ISGMR}$ where the BCPM(v) method is used are in
good agreement with the experimental data for almost all of the
considered nuclei, which is not the case when the Brueckner EDF is
used.

\begin{table}[h]
\caption{\label{tab:table4} The values of $E_{ISGMR}$ calculated
using Brueckner and BCPM(v) density functionals and
Eqs.~(\ref{eq:1}), (\ref{eq:35}) and (\ref{eq:36}) and compared
with the experimental data (the corresponding references are given
in Table~\ref{tab:table2}).}
\begin{ruledtabular}
\begin{tabular}{cccccc}
Nuclei&$r_0$&$r_0A^{1/3}$&$E_{ISGMR}^{Brueckner}$&$E_{ISGMR}^{BCPM(v)}$&Exp.\\[1pt]
      &[fm] & [fm]              &[MeV]                  &[MeV]             &[MeV]\\
\hline\\[-8pt]
 $^{40}$Ca&1.205&4.12&16.29&20.54&$19.18\pm0.37$\\
 $^{42}$Ca&1.199&4.17&16.20&20.46&$19.7\pm0.1$\\
 $^{44}$Ca&1.193&4.21&16.04&20.37&$19.49\pm0.34$\\
 $^{46}$Ca&1.187&4.25&15.79&20.11&--\\
 $^{48}$Ca&1.182&4.30&15.50&19.93&$19.88\pm0.16$\\
 $^{54}$Fe&1.168&4.41&15.86&19.97&$19.66\pm0.37$\\
 $^{56}$Ni&1.164&4.45&15.79&19.86&$19.1\pm0.5$\\
 $^{58}$Ni&1.160&4.49&15.55&19.61&$18.43\pm0.15$\\
 $^{60}$Ni&1.157&4.53&15.28&19.35&$17.62\pm0.15$\\
 $^{68}$Ni&1.144&4.67&14.16&18.33&$21.1\pm1.9$\\
 $^{64}$Zn&1.150&4.60&14.85&18.82&$18.88\pm0.79$\\
 $^{68}$Zn&1.144&4.67&14.43&18.44&$16.6\pm0.17$\\
 $^{90}$Zr&1.120&5.02&13.57&17.14&$16.9\pm0.1$\\
 $^{92}$Zr&1.118&5.05&13.35&16.99&$16.5\pm0.1$\\
 $^{92}$Mo&1.118&5.05&13.57&17.10&$16.6\pm0.1$\\
 $^{94}$Mo&1.116&5.07&13.39&16.95&$16.4\pm0.2$\\
 $^{96}$Mo&1.114&5.10&13.18&16.77&$16.3\pm0.2$\\
$^{106}$Cd&1.107&5.24&12.94&16.36&$16.27\pm0.09$\\
$^{110}$Cd&1.105&5.29&12.67&16.11&$15.94\pm0.07$\\
$^{112}$Cd&1.103&5.32&12.52&16.00&$15.80\pm0.05$\\
$^{114}$Cd&1.102&5.34&12.38&15.87&$15.61\pm0.08$\\
$^{116}$Cd&1.101&5.37&12.21&15.71&$15.44\pm0.06$\\
$^{100}$Sn&1.111&5.16&13.37&16.73&--\\
$^{112}$Sn&1.103&5.32&12.72&16.11&$16.2\pm0.1$\\
$^{114}$Sn&1.102&5.34&12.60&15.99&$16.1\pm0.1$\\
$^{116}$Sn&1.101&5.37&12.45&15.85&$15.8\pm0.1$\\
$^{118}$Sn&1.100&5.40&12.31&15.75&$15.8\pm0.1$\\
$^{120}$Sn&1.099&5.42&12.18&15.64&$15.7\pm0.1$\\
$^{122}$Sn&1.098&5.45&12.04&15.54&$15.4\pm0.1$\\
$^{124}$Sn&1.097&5.47&11.91&15.43&$15.3\pm0.1$\\
$^{132}$Sn&1.093&5.57&11.39&15.04&--\\
$^{204}$Pb&1.069&6.29&10.10&13.21&$13.98$\\
$^{206}$Pb&1.069&6.31&10.03&13.17&$13.94$\\
$^{208}$Pb&1.068&6.33&9.95&13.11&$13.96\pm0.2$
\end{tabular}
\end{ruledtabular}
\end{table}

Here we turn again to the open problem, namely which one of the
two procedures based on the use of Eq.~(\ref{eq:1}) or
Eq.~(\ref{eq:2}) has to be applied in the theoretical
calculations. In our opinion, the definitions in Eq.~(\ref{eq:1})
and Eq.~(\ref{eq:2}) reflect the existing difference between the
two usually used meanings of the so-called quantity "radius" of
the nucleus. One of them is the rms radius [whose definition is
well known, see, e.g., Eq.~(\ref{eq:32})] and another one is taken
usually to be the so-called "half-radius" as $R_{1/2}=r_{0}
A^{1/3}$, defined by $\rho(R_{1/2})=\rho(r=0)/2$. In the present
work we make a methodical semi-empirical analysis of the problem.
We use in both cases [Eqs.~(\ref{eq:1}) and (\ref{eq:2})] the
HF+BCS density $\rho(r)$ and the corresponding weight function
$|F(x)|^2$ in order to calculate the nuclear incompressibility
$K^{A}$ for a given EDF. In this way, in the case of
Eq.~(\ref{eq:2}) the analysis is self-consistent, namely because
both quantities $K^{A}$ and the rms radius are calculated by the
HF+BCS nuclear density. In the case of Eq.~(\ref{eq:1}) we use
again the same HF+BCS density and $K^{A}$. However, we note that
in the majority of nuclei it is impossible to measure the central
nuclear density $\rho(r=0)$ and, consequently, to obtain the
half-radius $R_{1/2}$ that takes part in Eq.~(\ref{eq:1}). Due to
this difficulty we introduced our semi-empirical approach given
above and obtained the parameterized form of $r_{0}$
[Eq.~(\ref{eq:35})] and $R_{1/2}=r_{0}A^{1/3}$ [Eq.~(\ref{eq:36})]
(including the "surface" term). The existing difference between
the rms radius and the $R_{1/2}$ leads to the difference between
the energies $E_{ISGMR}$ calculated by using Eq.~(\ref{eq:1}) or
Eq.~(\ref{eq:2}). In this way, following our method to obtain
these energies, we think that it is possible to shed more light,
at least approximately, on the existing problem. Having in mind
the lack of precise data for the neutron (and matter) rms radii
for the majority of the considered nuclei and the comparison of
the results shown in Table~\ref{tab:table2} and
Table~\ref{tab:table4}, in our opinion, the use of
Eq.~(\ref{eq:1}) with $r_{0}$ obtained by a procedure given above
[Eqs.~(\ref{eq:35}) and (\ref{eq:36}), see also Fig.~\ref{fig2}]
is more preferable.

We should emphasize also that the comparison of the results for
$E_{ISGMR}$ obtained using Eq.~(\ref{eq:1}) and both methods for
EDF [the Brueckner and BCPM(v) ones] with the data given in
Table~\ref{tab:table4} shows the advantage of the BCPM(v) method.

\section{Summary and conclusions}
\label{sec:conclusions}

In summary, we have performed a systematic analysis of the
isoscalar giant monopole resonance excitation energies in a wide
spectrum of nuclei within the microscopic self-consistent Skyrme
HF+BCS method with Skyrme SLy4 interaction and pairing
correlations, as well as the coherent density fluctuation model.
The method of calculations includes three steps. The first one is
to determine the incompressibility of infinite nuclear matter
[Eq.~(\ref{eq:15})]. For this purpose, for the potentia part of
the EDF we use that one of the Brueckner EDF and the one from the
EDF of Barcelona-Catania-Paris-Madrid [BCPM(v)]. The second step
includes calculations of the necessary incompressibility of finite
nuclei within the CDFM scheme averaging the incompressibility of
nuclear matter by the weight function $|F(x)|^{2}$
[Eq.~(\ref{eq:14})] that is related to the nucleon density
distribution [Eq.~(\ref{eq:9})]. The CDFM provides this
possibility being based on the delta-function limit of the
generator coordinate method and as extension of the Fermi gas
model. In this way, the CDFM allows one to obtain the quantities
of finite nuclei (such as the symmetry energy, the
incompressibility and others) on the base of the corresponding
ones for nuclear matter. As a third step, we perform calculations
of the centroid energies of the ISGMR based on the two
definitions: the one that uses the radial parameter of the density
distributions [Eq.~(\ref{eq:1})] and the other, in which the
excitation energy is expressed through the mean square mass radius
of the nucleus in the ground state [Eq.~(\ref{eq:2})]. The nuclear
incompressibility $K^{A}$ in both cases is calculated by using
Eq.~(\ref{eq:14}).

The results for the centroid energy deduced from Eq.~(\ref{eq:2})
in the case of the Brueckner EDF are in good agreement with the
data for the lighter isotopes of Ca, Fe, and Ni and acceptable for
those of Zn, Mo, and Cd. In the case of BCMP(v) EDF they are
comparable for some Pb isotopes and, in particular, for $^{68}$Ni
nucleus. The excitation energy of ISGMR in $^{68}$Ni is located at
higher energy (21.1 MeV) for the Ni isotopic chain due to the
large fragmentation of the isoscalar monopole strength
\cite{Vandebrouck2014,Vandebrouck2015}. Nevertheless, our value of
21.74 MeV reproduces very well the experimental one.

We have analyzed the problems related to the neutron density
distributions which cannot be precisely obtained in many cases, as
well as the corresponding rms radii. This point is of particular
importance for the neutron-rich nuclei, which are the majority of
the nuclei considered. Along this line, we preformed calculations
of the ISGMR energy in cases when data for the neutron skin
thickness are available. By using Eq.~(\ref{eq:2}) we found that
the results for the Sn isotopic chain ($A=112-124$), as well as
for the double-magic $^{48}$Ca and $^{208}$Pb nuclei recently
explored in CREX and PREX experiments, exhibit almost
non-sensitivity regarding the extracted mean square mass radius to
get values of the centroid energy.

By parametrization of the mass dependence of the radial parameter
in the range between $^{40}$Ca and $^{208}$Pb nuclei, we have
applied Eq.~(\ref{eq:1}) to calculate the ISGMR energies. A good
overall agreement with the experimental data for all considered
nuclei is achieved when using the BCPM(v) functional. In
conclusion, we would like to note that due the the lack of precise
data for the neutron (and matter) rms radii the use of
Eq.~(\ref{eq:1}) is more preferable in the theoretical
calculations. In addition, we have noted that the comparison of
the results for $E_{ISGMR}$ using Eq.~(\ref{eq:1}) and both EDF's
of Brueckner and BCMP(v) shows the advantage of the second one.
Finally, it is a hope that new data and refined theoretical
methods are necessary for the correct description of the breathing
modes in medium-heavy nuclei, more specifically the placement of
the ISGMR centroid energy.

\begin{acknowledgments}
M.K.G. is thankful to Prof. Evgeni Kolomeitsev for the useful
discussion. I.C.D wishes to acknowledge partial financial support
from the University of Mount Olive Professional Development Fund.
\end{acknowledgments}

\appendix\section{}\label{appA}

(1) The quantities $S(\rho_0(x))$ [Eq.~(\ref{eq:12})],
$K_{NM}(\rho_0(x))$ [Eq.~(\ref{eq:16})], $K_{sym}(\rho_0(x))$
[Eq.~(\ref{eq:19})], $L_{sym}(\rho_0(x))$ [Eq.~(\ref{eq:20})],
$Q(\rho_0(x))$ [Eq.~(\ref{eq:21})], and $B(\rho_0(x))$
[Eqs.~(\ref{eq:22}) and (\ref{eq:23})] obtained in the present
work in the case of the Brueckner EDF (Refs.~\cite{Brueckner68,
Brueckner69, Brueckner68a, Brueckner70}, see also e.g.
Refs.~\cite{Gaidarov2011, Danchev2020}) are as follows:
\begin{multline}\label{eq:A1}
S(\rho_0(x)) = \frac{5}{9}C \rho_0^{2/3}(x) + b_4 \rho_0(x) + b_5 \rho_0^{4/3}(x) \\+ b_6 \rho_0^{5/3}(x),
\end{multline}
\begin{multline}\label{eq:A2}
K_{NM} (\rho_0(x)) = -2C \rho_0^{2/3}(x) + 4 b_2 \rho_0^{4/3}(x) \\+10 b_3 \rho_0^{5/3}(x),
\end{multline}
\begin{multline}\label{eq:A3}
K_{sym} (\rho_0(x)) = -\frac{10}{9} C \rho_0^{2/3}(x) + 4 b_5 \rho_0^{4/3}(x) \\+ 10 b_6 \rho_0^{5/3}(x),
\end{multline}
\begin{multline}\label{eq:A4}
L_{sym} (\rho_0(x)) = \frac{10}{9} C \rho_0^{2/3}(x) + 3 b_4 \rho_0(x)\\
+ 4 b_5 \rho_0^{4/3}(x) + 5 b_6 \rho_0^{5/3}(x),
\end{multline}
\begin{align}\label{eq:A5}
Q (\rho_0(x)) = 8 C \rho_0^{2/3}(x) - 8 b_2 \rho_0^{4/3}(x) - 10 b_3 \rho_0^{5/3}(x).
\end{align}
The quantity $B(\rho_0(x))$ follows from Eqs.~(\ref{eq:22}) and (\ref{eq:23}).

In the equations given above:
\begin{align}\label{eq:A6}
  C=\frac{3}{5}\Big(\frac{h^2 c^2}{2m}\Big)\Big(\frac{3\pi^2}{2}\Big)^{2/3}\cong75.05~[\text{MeV fm$^3$}];
\end{align}
\begin{eqnarray}\label{eq:A7}
  b_1&=&-741.28,~b_2=1179.89,\notag\\
  b_3&=&-467.54,~b_4=148.26,\\
  b_5&=&372.84,~b_6=-769.57.\notag
\end{eqnarray}

(2) The quantities of nuclear matter mentioned in the previous
point~(1) but obtained in the case of the BCPM(v) EDF are as
follows:
\begin{multline}\label{eq:A8}
S(\rho_0(x)) = \frac{5}{9} C \rho_0^{2/3}(x) + \left\{- \sum\limits_{n=1}^{5}{a_n \Big(\frac{\rho_0(x)}{\rho_{\infty}}\Big)^{n}}\right.\\
\left. + \sum\limits_{n=1}^{5}{b_n \Big(\frac{\rho_0(x)}{\rho_{0n}}\Big)^{n}}\right\},
\end{multline}
\begin{multline}\label{eq:A9}
K_{NM}(\rho_0(x)) = -2C\rho_0^{2/3}(x) \\+ 9 \sum\limits_{n=1}^{5}{a_n n (n-1)\Big(\frac{\rho_0(x)}{\rho_\infty}\Big)^{n}}
\end{multline}
\begin{multline}\label{eq:A10}
K_{sym}(\rho_0(x)) = -\frac{10}{9}C\rho^{2/3}_0(x)\\ +
9 \left\{ - \sum\limits_{n=1}^{5}{a_n n (n-1)\Big(\frac{\rho_0(x)}{\rho_{\infty}}\Big)^{n}}\right. \\\left.+ \sum\limits_{n=1}^{5}{b_n n (n-1)\Big(\frac{\rho_0(x)}{\rho_{0n}}\Big)^{n}} \right\}
\end{multline}
\begin{multline}\label{eq:A11}
L_{sym}(\rho_0(x)) =  \frac{10}{9}C\rho^{2/3}_0(x) - 3
\sum\limits_{n=1}^{5}{a_n n
\Big(\frac{\rho_0(x)}{\rho_\infty}\Big)^{n}}\\ +
3\sum\limits_{n=1}^{5}{b_n n
\Big(\frac{\rho_0(x)}{\rho_{0n}}\Big)^{n}},
\end{multline}
\begin{multline}\label{eq:A12}
Q(\rho_0(x)) =  8C\rho^{2/3}_0(x)\\ +27 \sum\limits_{n=1}^{5}{a_n
n (n-1) (n-2) \Big(\frac{\rho_0(x)}{\rho_\infty}\Big)^{n}}.
\end{multline}
The quantity $B(\rho_0(x))$ follows from Eqs.~(\ref{eq:22}) and (\ref{eq:23}).

\end{document}